\documentclass[aps, prd, preprint, preprintnumbers, showpacs, showkeys, amsfonts, nofootinbib, 
floatfix,superscriptaddress] {revtex4-1}
\usepackage{graphicx,graphics}
\usepackage{amssymb}
\usepackage{amsthm}
\usepackage{pstricks}
\usepackage{color}
\usepackage{slashed}
\usepackage{newlfont}
\usepackage{ulem}
\usepackage{array}
\usepackage{float}
\usepackage{amsmath,amsfonts}
\usepackage{subfigure}
\usepackage{slashed}
\def\beq{\begin{equation}}
\def\eeq{\end{equation}}

\def\bt{\begin{table}}
\def\et{\end{table}}
\def\bc{\begin{center}}
\def\ec{\end{center}}
\def\bi{\begin{itemize}}
\def\ei{\end{itemize}}
\def\bea{\begin{eqnarray}}
\def\eea{\end{eqnarray}}
\def\beas{\begin{eqnarray*}}
\def\eeas{\end{eqnarray*}}

\def\nn{\nonumber}
\def\NPB#1,{{\rm Nucl.\ Phys.\ B }{\bf #1},}
\def\PLB#1,{{\rm Phys.\ Lett.\ B }{\bf #1},}
\def\EPJC#1,{{\rm Eur.\ Phys.\ Jour.\ C }{\bf #1},}
\def\PRD#1,{{\rm Phys.\ Rev.\ D }{\bf #1},}
\def\PRL#1,{{\rm Phys.\ Rev.\ Lett.\ }{\bf #1},}
\def\MPLA#1,{{\rm Mod.\ Phys.\ Lett.\ A }{\bf #1},}
\def\JPG#1,{{\rm J.\ Phys.\ G }{\bf #1},}
\def\CTP#1,{{\rm Commun.\ Theor.\ Phys.\ }{\bf #1},}
\def\JHEP#1,{{\rm JHEP \ }{\bf #1},}
\def\NPPS#1,{{\rm Nucl.\ Phys.\ Proc.\ Suppl.\ }{\bf #1},}
\def\CPC#1,{{\rm Computl.\ Phys.\ Commun.\ }{\bf #1},}
\def\CPL#1,{{\rm Chin.\ Phys.\ Lett. }{\bf #1},}
\def\APPB#1,{{\rm Acta\ Phys.\ Polon.\ B }{\bf #1},}
\def\AHEP#1,{{\rm Adv.\ High Energy\ Phys.\  }{\bf #1},}
\def\PR#1,{{\rm Phys.\ Rev.\ }{\bf #1},}
\def\sq2{\sqrt{2}}

\def\G{\Gamma}
\def\l{\lambda}
\def\s{\sigma}

\begin{document}
\preprint {RECAPP-HRI-2014-009; \, OSU-HEP-14-05}

\title{\large {\bf New signals for singlet Higgs and vector-like quarks at the LHC} }

\author{Durmu\c{s} Karabacak}
\email{Electronic address: durmus.karabacak@.okstate.edu}
\affiliation{Department of Physics and Oklahoma Center for High Energy Physics,
Oklahoma State University, Stillwater OK 74078--3072, USA}
\author{S. Nandi}
\email{Electronic address: s.nandi@okstate.edu}
\affiliation{Department of Physics and Oklahoma Center for High Energy Physics,
Oklahoma State University, Stillwater OK 74078--3072, USA}
\author{Santosh Kumar Rai}
\email{Electronic address: skrai@hri.res.in}
\affiliation{Regional Centre for Accelerator-based Particle Physics, 
Harish-Chandra Research Institute,  Chhatnag Road, Jhusi, Allahabad 211019, 
India}

\begin{abstract}
We consider an extension of the Standard Model involving a singlet Higgs and down type vector-like quarks in the light of the current LHC Higgs data. For a good range of the parameters of the Higgs potential, and a mass range for the heavy vector-like quark,  we find that the singlet heavy Higgs arising from the production and decay of the vector-like quarks give rise to (2b~4t) signal. The subsequent decay of the top quarks to $b W^{+}$ give rise to a final state with six b quarks, two same-sign charged leptons and missing transverse momenta with observable cross-sections  at the 14 TeV run of the Large Hadron Collider. The Standard Model background for such a final state is practically negligible.
\end{abstract}

\maketitle

\section{\label{sec:intro}Introduction}
The discovery of the Higgs-like boson at the CERN Large Hadron Collider (LHC)
is certainly a great success of the Standard Model (SM) \cite{Aad:2012tfa}\cite{Chatrchyan:2012ufa}. Though not completely established, this particle looks very much like the SM Higgs boson. However, ATLAS Collaboration \cite{ATLAS_diphoton} results for the $\gamma \gamma$ signal both for the production in the gluon-gluon fusion mode as well as from the associated production with the vector boson gives significant enhancement compared to the SM prediction. The CMS Collaboration results for the same modes gives modest suppression in both channels. A combined fit \cite{Belanger:2013xza} (admittedly by the theoreticians) fitting all channel data from the LHC as well as Tevatron gives again significant enhancement in the $\gamma \gamma$ channel. It is not clear at this time if this is a signal of new physics or not. However, any new physics which can fit the data better is worth exploring. On the theoretical vein, there is no fundamental reason why there should be only one SM Higgs boson, or only chiral fermions. In fact, most extensions of the SM includes more Higgs bosons, and also non-chiral fermions. Any reasonably motivated model which gives good agreement with the Higgs data, as well as predict new physics that can be tested at the LHC is worth exploring. It is in this spirit, we consider a model which extends the SM by including down type vector-like quark ($D$) \cite{vlq_theory}
and add a real singlet Higgs boson ($S$) \cite{singlet} to the scalar sector. In this model, we are able to fit the LHC data from observed Higgs final states, especially the $\gamma \gamma$ channel better because of the additional contribution in the loop coming from a new colored particle. The down type vector-like quark can be pair produced with good enough rates at the LHC. The heavy quark in our case will decay dominantly into the b-quark and the scalar singlet Higgs. This scalar Higgs then dominantly decays to $t \bar{t}$, and $t$ decays dominantly to $b W^{+}$. Thus from the pair production of the of $D \bar{D}$, and from the subsequent decay of the two top quarks in the leptonic mode (via $b W^{+}$), we get the final state with six $b$ (three $b$ and three $\bar{b}$), two same sign charged leptons ($e$ or $\mu$) and $\slashed{E}_T$. 
We find that for a wide range of the parameter space involving coefficients in the Higgs potential, and for a wide range of mass for the vector-like quark, such an exotic final state yields observable event rates at the 14 TeV run of LHC, even with a modest luminosity of $100~fb^{-1}$. 
The SM background for such a final state will be too suppressed and totally negligible. Thus the observation of any events for such a final state will be a signal for new physics beyond the SM. 

\section{The Model and the Formalism}\label{sec:model} 

The gauge symmetry of our model is the same as the SM, {\it viz.}, $SU(3)_c \times SU(2)_L \times U(1)_Y$. 
We extend the matter sector of the SM with an additional down type vector-like quark, $D$ and the scalar sector with a real scalar singlet field, $S$. The Lagrangian of our model is given by  
\beq\label{eq:lag}
\begin{split}
\mathcal{L}=~&\mathcal{L}_{SM}-\bar{D}(i\gamma^{\mu}D_{\mu}-M_{D})D-f_{D}\bar{D}DS,
\end{split}
\eeq	
where $M_D$ is the bare mass term for the vector-like quark (VLQ) while we have also added a gauge-singlet 
Yukawa interaction term for the vector like quark with the new scalar singlet whose coupling strength is given
by $f_D$.
The scalar potential is given by 
\beq\label{eq:potential}
\begin{split}
V(H,S) = & -\mu_{1}^2(H^{\dagger} H)-\mu_{2}^2S^2 +     \lambda_{1}(H^{\dagger} H)^2 + \lambda_{2} S^4 \\ 
& +\lambda_{3}(H^{\dagger} H )S^2 +\sigma_{1}S^3 +\sigma_{2}(H^{\dagger} H )S~.             
\end{split}
\eeq
where the parameters $\mu_1, \mu_2, \sigma_1$ and, $\sigma_2$ have mass dimensions. The 
electroweak (EW) symmetry is spontaneously broken when the neutral component of the Higgs doublet $H$ gets a {\it vacuum expectation value} (VEV). In the unitary gauge the shifted VEV's of the $H$ and $S$ can be 
written as
\begin{equation}\label{eq:vevs}
H =\frac{1}{\sqrt{2}}
\begin{pmatrix}
0  \\
v_{h}+h_{0} 
\end{pmatrix}
\quad
,~~~S =v_{s}+s_{0}
\end{equation}
where $v_{h}$ and $v_{s}$ are VEV's of corresponding scalar fields.
Note that the vector-like quark gets a bare mass as well as a mass from its Yukawa interaction with the singlet Higgs.

Minimizing the scalar potential, we get the following constraints among the parameters given by:
\beq\label{eq:potential-cond}
\begin{split}
&\mu_{1}^2 = \lambda_1 v_{h}^2 +\sigma_{2} v_s +\lambda_3 v_s^2, \hspace*{0.3in}
\mu_{2}^2 = \frac{\sigma_2 v_h^2}{4v_s}+\frac{\lambda _{3}}{2}v_h^2+\frac{3}{2}\sigma_{1}v_s+2 \lambda_{2}v_s^2,\\
&\lambda_{1} > 0,  \hspace*{0.3in} 
3\sigma_1 v_s+8\lambda_{2}v_s^{2}-\frac{\sigma_2 v_h^2}{2 v_s} > 0,  \hspace*{0.3in}
\sigma_{2}+2\lambda_{3}v_s > 0.
\end{split}
\eeq
In addition to the constraints in Eq. \ref{eq:potential-cond} one needs to assume $\lambda_{2}>0$ so that the potential is bounded from below for large values of the singlet field. The scalar mass squared matrix in $(h_{0},s_{0})$ basis is given by
\beq\label{eq:massm}
\mathcal{M}^2=
\begin{pmatrix}
2\lambda_{1} v_{h}^2 & v_{h}(2\lambda_{3}v_{s}+\sigma_{2})\\
v_{h}(2\lambda_{3}v_{s}+\sigma_{2}) & ~~8\lambda_{2}v_{s}^2+3\sigma_{1}v_{s}-\frac{\sigma_{2}v_{h}^2}{2v_{s}}
\end{pmatrix}.
\eeq
The fields $(h_0,s_0)$ can be expressed in terms of the physical fields $(h,s)$ as
\beq
\begin{split}
h_{0} =&~~ h\cos\beta + s\sin\beta\\
s_{0} =& -h\sin\beta+ s\cos\beta.
\end{split}
\eeq
The mixing angle $\beta$ is given by 
\beq
\text{tan}2\beta=\frac{2 \mathcal{M}^2_{12}}{\mathcal{M}^{2}_{22}-\mathcal{M}^{2}_{11}}~.
\eeq 
where $\mathcal{M}^{2}_{ij}$ is the $(i,j)^{th}$ element of $\mathcal{M}^2$ in 
Eq.~\ref{eq:massm}. The VLQ, in principle can also mix with other SM quarks and assuming that the vector-like heavy quark, D dominantly couples to only the b quark in its Yukawa interactions with the singlet $S$ and the Higgs doublet $H$, we can write the most general gauge invariant Yukawa interaction and mass terms,
that lead to a mixing of the VLQ with the $b$ quark, given by
\beq\label{eq:vlmixing}
\begin{split}
-\mathcal{L}_{bD}=& ~y_{b}\bar{Q}_{3L} b_{R} H + M_{D}\bar{D}_L D_R + f_{D}\bar{D}_L D_R S 
+ f_{Q H} \bar{Q}_{3L} D_{R} H \\  & ~ + Y_{bD}^{*}\bar{D}_{L} b_{R} S + M_{bD} \bar{D}_{L}b_{R}+\textit{h.c.}  \\
\end{split}
\eeq 
In Eq.~\ref{eq:vlmixing} $y_{b}, f_{D}, f_{QH}$, and $Y_{bD}$ are Yukawa couplings, while 
$ M_{bD}$ and $ M_{D}$ are bare mass terms in the Lagrangian. Using the above Lagrangian, we have calculated the mass eigenstates from the mixing matrix 
for the $b$ and $D$, and their left and right mixing angles ($\theta_L,\theta_R$) using bi-unitary transformations. 
Note that the mixing angles $\theta_L$ and $\theta_R$ are constrained by observables involving $b$ quarks, in interactions within the SM as well as the entries in the Cabibbo-Kobayashi-Maskawa (CKM) matrix, which we have used in our calculations.

\section{Phenomenology and signals at the LHC}

We now consider the final states that highlight a very interesting and unique signal arising from the pair productions of these vector-like $D$ quarks and their subsequent decays.

\begin{figure}[t!]
\centering
\includegraphics[width=4.8in,height=2.8in]{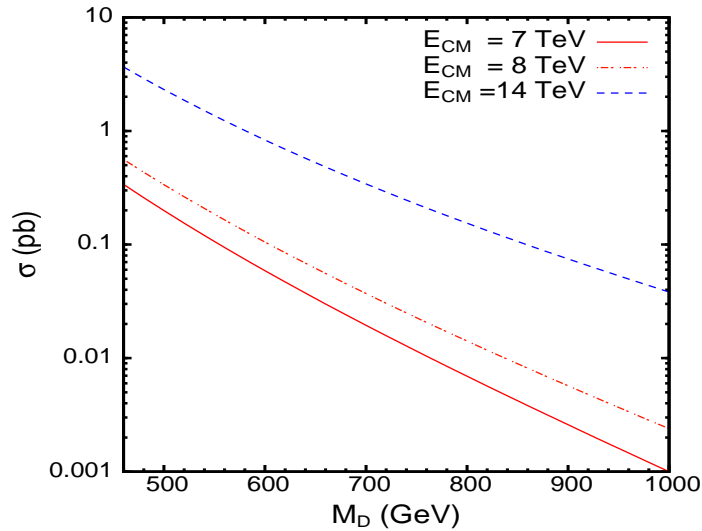}
\caption{Leading order production cross-section for 
$p p \rightarrow D \bar{D}$ at the LHC as a function of 
VLQ mass $M_{D}$, at center-of-mass energies, 
$E_{CM} = 7, 8$ and $14$ TeV. We have chosen the scale as 
$Q=M_{D}$, the mass of the heavy VLQ.}
\label{fig:cs} 
\end{figure}

In Fig.\ref{fig:cs} we plot the leading order (LO) production cross-section of $D \bar{D}$ at $7,~8,~\text{and}~14$ TeV center-of-mass energies at the LHC. The factorization scale $Q$ has been set to the mass of $D$, $M_{D}$, and we have used the {\tt CTEQ6$\ell$1}~\cite{Pumplin:2002vw} parton distribution function (PDF) set. The model is also implemented in {\tt CalCHEP}~\cite{Belyaev:2012qa} and the results have been found in good agreement with our parton level Monte Carlo generator.
The production cross-section at 8 TeV center-of-mass energy is above $100$ fb at mass $M_D=600$ GeV and drops below $10$ fb after $M_D=800$ GeV. There already exist search limits by both the CMS \cite{CMS_vlq} and ATLAS \cite{ATLAS_vlq} Collaborations on such exotic quarks. However, the search limits crucially depend on how these exotic quarks decay, and most of the limits assume the decays of the bottom-like exotics in to the $tW$ channel with 100\% branching probability (and/or only considering $ bh $, $ bZ $ and $ tW $ decay channels). In this model, the mixing of the VLQ quark, $D$ with the SM down-type quarks will dictate its decay properties. In our case, by construction its mixing is expected to be dominant with the $ b $-quark. The mixing angles, $\theta_{L}$ and $\theta_{R}$, also allow two decay channels, namely $D \to t ~W^-$ and $D \to b ~Z$. Both is a direct consequence of the extra mixing in the quark sector. The singlet scalar field also mixes with the SM Higgs doublet. This mixing $\beta$ induces two more decay channels for the VLQ, namely $D \to b ~s$ and $D \to b ~h$. 

\begin{figure}[ht!]
\includegraphics[height=3.1in,angle = 0,width=6.0in]{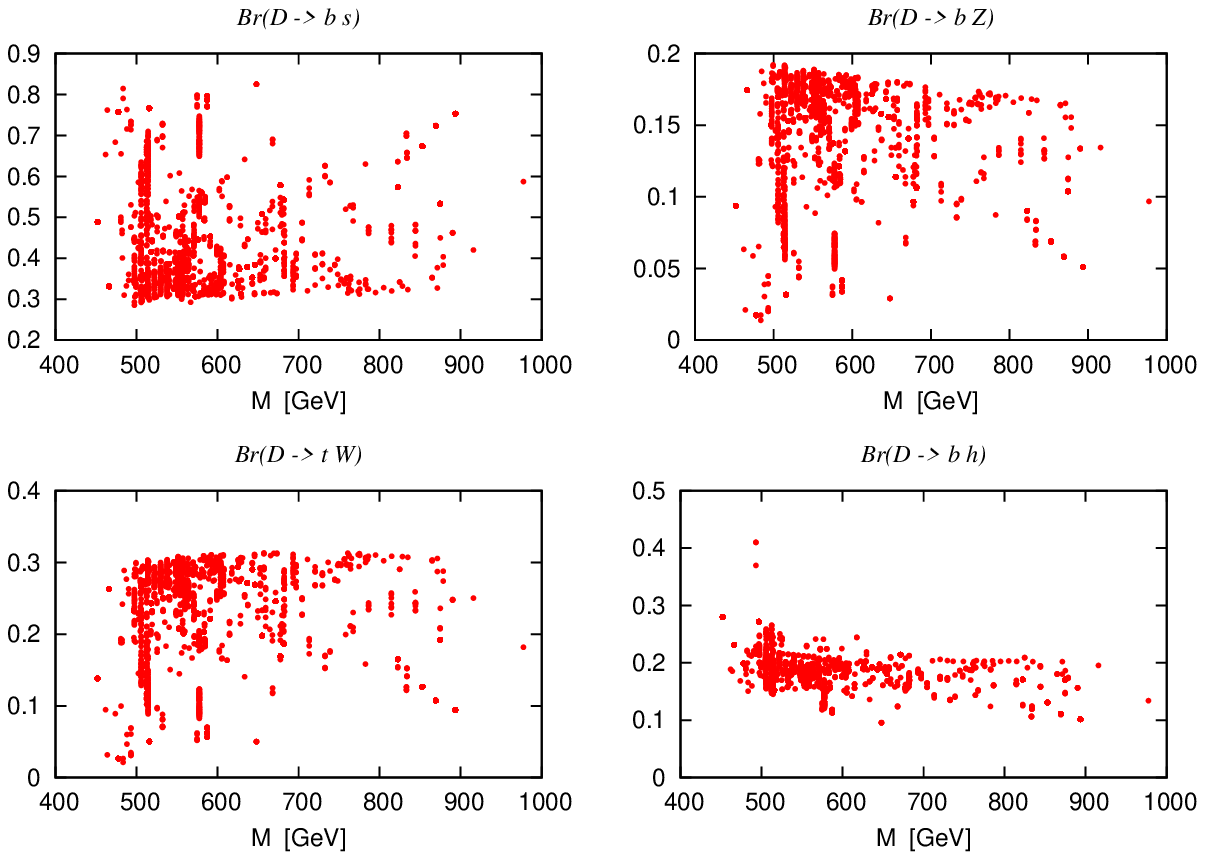}
\caption{ The branching ratio of VLQ versus its mass plot for four decay channels resulting from the parameter scan. From top left corner to right bottom corner plots correspond to $D \to b s, b Z, t W$ and $b h$ decays, respectively. Note that unity of sum of the branching ratios is satisfied.} 
\label{fig:D_branching}
\end{figure}

\begin{figure}[t!]
\includegraphics[height=3.1in,angle = 0,width=6.0in]{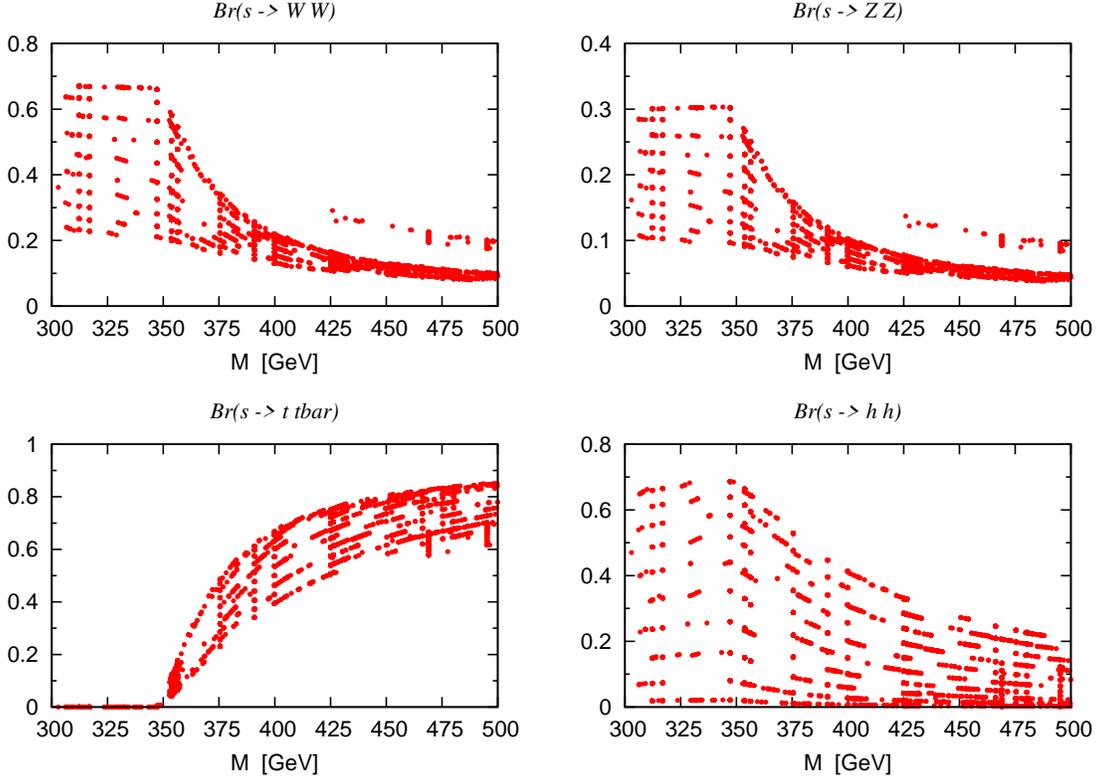}
\caption{The branching ratio of the heavy Higgs $(s)$ versus its mass for four decay channels resulting from the parameter scan. From top left corner to right bottom corner plots correspond to $s \to W^+ W^-,~Z Z,~t \bar{t}$  and $h h$ decays, respectively. Note that unity of sum of the branching ratios is satisfied.}
\label{fig:S_branching}
\end{figure}

To highlight the signal in our model we scanned over values for the free parameters in our model. The range of the parameter space scanned are shown in Table I where we have listed the parameters appearing in the scalar potential and in Table II, where we list the free parameters that constitute the Yukawa and mass terms for the VLQ and $b$ quark. 
\begin{table}[t!]
    \begin{tabular}{c|c|c|c|c|cc} \hline \hline
    $\l_1$ & $\l_2$ & $\l_3$ & $\s_1$(GeV) & $\s_2$(GeV) & $v_s$(GeV) \\
    \hline \hline
    $[0.1,1.1]$ & $[0.1,1.1]$ & $[0.1,1.1]$ & $[-500,500]$ & $[-500,500]$ & $[100,500]$ & \\ \hline 
    \end{tabular}    
    \caption{Illustrating the range over which the free parameters of the scalar sector are varied. The doublet VEV $v_h$ is fixed at $246$ GeV.}\label{tb:scan_scalar}
\end{table}
\begin{table}[b!]
    \begin{tabular}{c|c|c|c|c|cc} \hline  \hline
    $f_D$ & $y_b$ & $Y_{bD}$ & $f_{QH}$ & $M_{bD}$(GeV) & $M_D$(GeV) \\
    \hline  \hline
    $[-1,1]$ & $(0,1]$ & $[-1,1]$ & $[-1,1]$ & $[-500,500]$ & $[-500,500]$ & \\ \hline
    \end{tabular}
    \caption{Illustrating the range over which the free parameters associated with the Yukawa and mass terms in the Lagrangian (involving $b$ quark and the VLQ) are varied.}\label{tb:scan_quark}
\end{table}
In the numerical scan over the free parameters in our model, we demanded that the modified 
signal strengths ratio 
\beq
\mu_{XX} = \frac{\sigma^{NEW}(g g \to h)\times\G^{NEW}(h \to X X)}{\sigma^{SM}(g g \to h)\times\G^{SM}(h \to X X)}
\eeq 
would be within 10\% of the SM expectations, assuming a quite conservative restriction when compared 
to the actual signal strengths as observed by ATLAS and CMS Collaborations.
We find that a wide range of the parameter space satisfies all the observed Higgs data at the LHC,  as well as other experimental constraints coming from interactions involving couplings such as $Zb\bar{b}$, $V_{CKM}$, and also direct search limits on heavy particle productions at accelerator experiments like LEP, Tevatron and LHC. From this, we found a significant range of the parameter space where the decay $D\rightarrow b s$ dominates, where $s$ is the heavy mass eigenstate coming dominantly from the singlet scalar. Note that such a decay channel has not been considered in the experimental searches, and therefore would weaken the existing constraints on VLQ mass. We present the  scattered branching ratios for 
\begin{align}
D \to b ~h,~~b ~s,~~t ~W,~~b ~Z 
\end{align}
 versus mass plot resulting from this parameter scan in Fig.~\ref{fig:D_branching}. 
To analyze the signal we have chosen two set of 
parameters which we treat as benchmarks as shown in Table~\ref{tb:benchmarks}. These 
points were then used to calculate 
the branching ratios for different final states, estimates 
of which are given in Table~\ref{tab:branching}. It is also instructive to look at the branching 
probabilities for the heavy Higgs decay for the same range of parameters which we show 
in Fig.~\ref{fig:S_branching}.
This scattered branching ratio versus mass plot can also be obtained from the same parameter scan for the heavy Higgs particle. Although the scan is over all decay channels for the heavy Higgs we only present the significant decays in the plot. In Fig.~\ref{fig:S_branching} we can see that the heavy Higgs will dominantly decay to two massive vector bosons and a pair of light Higgs until the 
decay to $t \bar{t}$ is kinematically allowed. What is new for the heavy Higgs is that there is a region of the parameter space in which one can suppress decays to vector bosons and enhance decay to light Higgs pair until $t\bar{t}$ threshold is reached. This feature is due to the fact that the couplings between two scalars can be tuned so that the decay to two light Higgses can be enhanced. The decay to $t\bar{t}$ starts when the heavy Higgs mass is around $350$ GeV and quickly dominates over the other decays with increasing mass. In the proceeding section we discuss the phenomenology of VLQ and the heavy Higgs with the benchmark points listed in Table~\ref{tb:benchmarks}.

\begin{table}[!h]
\begin{tabular}{|c||c|c|}
\hline {\bf Parameters} & $\mathcal{C}_1$ & $\mathcal{C}_2$\\ 
\hline $(\l_1,\l_2,\l_3)$ & (0.2, 0.2, 2.8) & (0.2, 0.1, 0.3) \\
\hline $(\sigma_1,\sigma_2,v_S)$  & (200, -400, 100) GeV & (100, -50, 360) GeV \\ 
\hline $y_b,f_D,f_{QH},Y_{Db}$ & 0.08, -0.5, 0.04, 0.16 & 0.24, -0.5, 0.04, 0.84 \\
           $(M_{bD},M_D)$ &  (450, 500) GeV &  (500, 400) GeV \\
\hline $m_h$  & 125.15 GeV & 125.9 GeV \\
\hline $M_s$  & 453.4 GeV & 353.6 GeV \\
\hline $M_{D}^{phys.}$  & 648 GeV & 853 GeV \\ \hline
\end{tabular}
\caption{Representative points in the model parameter space and
the relevant mass spectrum used in the analysis.}
\label{tb:benchmarks}
\end{table} 
\begin{table}[!t]
\begin{tabular}{|c||c|c|}
\hline {\bf Decay process} & $B(X \to Y ~Z)$ ~for~$\mathcal{C}_1$ & $B(X \to Y ~Z)$ ~for~$\mathcal{C}_2$\\ 
\hline $D \to b ~s$ & $0.825$ & $ 0.704 $ \\
\hline $D \to b ~h$  & $9.5 \times 10^{-2}$  & $0.105$  \\ 
\hline $D \to t ~W$ & $5.0 \times 10^{-2}$ & $0.122$ \\
\hline $D \to b~Z $  & $2.9 \times 10^{-2}$ & $6.6 \times 10^{-2}$  \\
\hline $s \to Z~Z$ &$4.9 \times 10^{-2}$ &$5.2 \times 10^{-2}$ \\
\hline $s \to t ~\bar{t}$ & $0.655$ & $0.832$ \\
\hline
\end{tabular}\label{tb:cs_times_br}
\caption{The branching ratios of VLQ, and that of the heavy Higgs used in the analysis, at two sets of parameter points. $M_D^{phys.}$ and $M_s$ are the physical mass for VLQ and the heavy Higgs resulted from the corresponding parameter sets.}
\label{tab:branching}
\end{table}
Finally  we are now ready to discuss the unique final state signal that arises in our model which is observable at the 14 TeV LHC with modest luminosity and which has negligible SM background. This follows from the pair production of $D \bar{D}$, and their subsequent decays following the decay chains given below:
\begin{align}
 p ~p \longrightarrow &  ~(D \to b~ s) \longrightarrow (s \to t~\bar{t})~b  \nonumber \\
  \hookrightarrow & ~(\bar{D} \to \bar{b} ~s)  \longrightarrow (s \to t~\bar{t})~\bar{b}    \\
  \hookrightarrow & ~t\bar{t}t\bar{t}+2b. \nn 
\end{align}
As can be seen from the parameter scan in Fig \ref{fig:D_branching}, for $D$ mass above $500$ GeV,  $D \to b ~s$ decay dominates over other decays. Thus $D \bar{D}$ pair productions give rise to ($bs$) ($\bar{b}s$). Also, as can be seen from Fig.\ref{fig:S_branching}, for $s$ mass above $450$ GeV,  the decay $s \to t~\bar{t}$ dominate over the other decay modes. Thus for a good range of parameter space allowed in our model, the final state from the $D\bar{D}$ production is 
$t\bar{t}t\bar{t} + 2b$ (where one $b$ is $\bar{b}$). The branching ratio for the $t$'s for the decays to $b W$ is essentially one. Thus the final state is $ 6 b + 2 W^{+} 2 W^{-}$. Now we consider either two  $W^{+}$ decay or two  $W^{-}$ decay leptonically to $e$ or $\mu$ plus neutrinos. The other two $W$'s decay hadronically or leptonically. Thus the resulting final state signal is six high $p_T$ $b$-jets, two high $p_T$ same sign charged leptons plus missing energy due to the neutrinos (where we do not trigger on the jets or charged leptons coming from the decays of the other $W's$). Let us now calculate this signal using our production cross-sections for the two benchmark points $\mathcal{C}_1$ and $\mathcal{C}_2$ given in given in Table \ref{tab:branching}. For example, $\mathcal{C}_1$ results in $M_D = 648$ GeV which has  $535~fb$ cross-section at $\sqrt{s} = 14$ TeV LHC. This cross-section when multiplied with the relevant branching ratios given in Table \ref{tab:branching}, $(B(D \to b~s)\times B(s \to t~\bar{t}))^2$, gives $156~fb$. This need to be multiplied by the leptonic branching ratios of the two same sign $W$'s decaying to $e$ or $\mu$ plus neutrinos, which is $\simeq 0.2$. Finally we have to multiply by the b-tagging efficiency \cite{btag} of each of the six $b$'s which is $\simeq 0.7$ for high $p_T$ $b$. Thus the resulting cross-section for the final state with six $b$'s and two same sign charged leptons is $\simeq 2 \times (156 ~fb) \times  (0.7)^6  \times (0.2)^2$  giving a value of $\simeq 1.4 ~fb$. Note that we have considered both final states with $l^+ l^+$ and $l^- l^-$. Thus with a modest $100~fb^{-1}$ luminosity, we expect $\simeq 140$ such events.  For our benchmark point $\mathcal{C}_2$, similar calculation yields $\simeq 40$ events. The SM background for this final state is negligible. Note that what we have presented here as an estimate of the signal events is just a crude estimate to highlight that the signal events are not negligible. However, as one would expect, after putting some basic acceptance cuts required to trigger on the different final states, the rates would be smaller. Even then, we do not expect the suppression to be more than 40--50\% of the estimated event rates and this still gives us significantly large and observable event rate for the signal, in the absence of any SM background.

\section{Conclusions} 

We have proposed a simple extension of the SM by extending the Higgs sector with a real singlet Higgs ($S$), and the matter sector with a down type vector-like quark ($D$). We have scanned the parameter space of the extended Higgs sector and the extended fermionic sector. There is a good range of the parameter space where all the experimental data from LEP, Tevatron and the LHC (including all the data of the measured cross-sections time the branching ratios of the observed 125 GeV Higgs) can be satisfied.
In this allowed parameter space, we found a wide range in which a unique final state with 6 b and two same sign charged leptons plus missing energy which will be well observable in the upcoming runs of the LHC, even with a modest luminosity of $100~fb^{-1}$. The SM background for such a final state is negligible and thus any observation of such a final state will be a clear signal of new physics beyond the standard model.

\begin{acknowledgments}
This research was supported in part by the United States Department of Energy under grant Number de-sc0010108. The work of  S.K.R. was partially supported by funding available from the Department of Atomic Energy, Government of India, for the Regional 
Centre for Accelerator-based Particle Physics, Harish-Chandra Research Institute.
\end{acknowledgments}

\end{document}